\documentclass[11pt,preprint]{aastex}

\newcommand{\hcccn}{HC$_3$N}

\newcommand{\kms}{km s$^{-1}$}

\shorttitle{Observations of \hcccn}
 \shortauthors{Li et al.}

\begin{document}

\title{TMRT Observations of Carbon-chain molecules in Serpens South 1A}

\author{Juan Li\altaffilmark{1, 2},
Zhi-Qiang Shen\altaffilmark{1, 2}, Junzhi Wang\altaffilmark{1, 2}, Xi Chen\altaffilmark{1, 2}, Ya-Jun Wu\altaffilmark{1, 2}, Rong-Bing Zhao\altaffilmark{1, 2}, Jin-Qing Wang\altaffilmark{1, 2}, Xiu-Ting Zuo\altaffilmark{1, 2}, Qing-Yuan Fan\altaffilmark{1, 2}, Xiao-Yu Hong\altaffilmark{1, 2}, Dong-Rong Jiang\altaffilmark{1, 2}, Bin Li\altaffilmark{1, 2},
Shi-Guang Liang\altaffilmark{1, 2}, Quan-Bao Ling\altaffilmark{1, 2}, Qing-Hui Liu\altaffilmark{1, 2}, Zhi-Han Qian\altaffilmark{1, 2}, Xiu-Zhong Zhang\altaffilmark{1, 2}, Wei-Ye Zhong\altaffilmark{1, 2}, Shu-Hua Ye\altaffilmark{1, 2}
}

\altaffiltext{1}{Department of Radio Science and Technology, Shanghai Astronomical observatory, 80 Nandan RD, Shanghai 200030, China; lijuan@shao.ac.cn}

\altaffiltext{2}{Key Laboratory of Radio Astronomy, Chinese Academy of Sciences, China}

\begin{abstract}

We report Shanghai Tian Ma Radio Telescope detections of several long carbon-chain molecules at C and Ku band, including HC$_3$N, HC$_5$N, HC$_7$N, HC$_9$N, C$_3$S, C$_6$H and C$_8$H toward the starless cloud Serpens South 1a. We detected some transitions (HC$_9$N $\emph{J}$=13-12 $\emph{F}$=12-11 and $\emph{F}$=14-13, H$^{13}$CCCN $\emph{J}$=2-1 $\emph{F}$=1-0 and $\emph{F}$=1-1, HC$^{13}$CCN $\emph{J}$=2-1 $\emph{F}$=2-2, $\emph{F}$=1-0 and $\emph{F}$=1-1, HCC$^{13}$CN $\emph{J}$=2-1 $\emph{F}$=1-0 and $\emph{F}$=1-1) and resolved some hyperfine components (HC$_5$N $\emph{J}$=6-5 $\emph{F}$=5-4, H$^{13}$CCCN $\emph{J}$=2-1 $\emph{F}$=2-1) for the first time in the interstellar medium. The column densities of these carbon-chain molecules in a range of 10$^{12}$-10$^{13}$ cm$^{-2}$ are comparable to two carbon-chain molecule rich sources, TMC-1 and Lupus-1A. The abundance ratios are 1.00:(1.11$\pm$0.15):(1.47$\pm$0.18) for [H$^{13}$CCCN]:[HC$^{13}$CCN]:[HCC$^{13}$CN]. This result implies that the $^{13}$C isotope is also concentrated in the carbon atom adjacent to the nitrogen atom in HC$_3$N in Serpens south 1a, which is similar to TMC-1. The [HC$_3$N]/[H$^{13}$CCCN] ratio of 78$\pm$9, the [HC$_3$N]/[HC$^{13}$CCN] ratio of 70$\pm$8, and the [HC$_3$N]/[HCC$^{13}$CN] ratio of 53$\pm$4 are also comparable to those in TMC-1. In any case, Serpens South 1a proves a testing ground for understanding carbon-chain chemistry.

\end{abstract}


\section{Introduction}
\qquad

Linear carbon-chain molecules like C$_n$H, HC$_{2n+1}$N, and C$_n$S are observed to be abundant in cold dense clouds (Bell et al. 1997; Kaifu et al. 2004; Kalenskii et al. 2004; Langston \& Turner 2007; Sakai et al. 2010a), in the circumstellar envelopes of carbon-rich Asymptotic Giant Branch (AGB) stars (Winnewisser \& Walmsley 1978; Cernicharo \& Guelin 1996; Guelin et al. 1997; Gong et al. 2015a), in massive star-forming regions (e.g. Watanabe et al. 2015; Gong et al. 2015b; Feng et al. 2015), in photo-dissociation regions (PDR; Pety et al. 2005; Gratier et al. 2013), and in diffuse clouds (e.g. Lucas \& Liszt 2000; Liszt et al. 2012). Detection of these molecules reveals that large organic molecules are forming in the interstellar medium (ISM). These species are proposed to be related to the formation and destruction of polyaromatic hydrocarbons (PAH; Henning \& Salama 1998; Tielens 2008) and may be carriers of some diffuse interstellar bands. These molecules are also considered to be an evolutionary indicator of prestellar cores in star formation studies (Suzuki et al. 1992; Li et al. 2012). The full characterization of carbon-chain molecules are thus regarded to be an important issue in astrochemistry (Sakai et al. 2010a).

However, studies of long carbon-chain molecules have been seriously limited by the small number of detections. To date, long carbon-chain molecules have been primarily seen towards several nearby, low-mass star-forming regions, and in the envelope of the carbon star (Bell et al. 1997; Sakai et al. 2010a; Cernicharo \& Guelin 1996; Guelin et al. 1997). For example, HC$_9$N has been detected in two nearby molecular clouds: Taurus (Broten et al. 1978; Sakai et al. 2008) and Lupus (Sakai et al. 2010a), while HC$_{11}$N was only detected in TMC-1 (Bell et al. 1997). As a consequence, many problems related to the formation and chemistry of carbon-chain molecules remain poorly understood.

The abundance decrement with increasing chain length is a key parameter for understanding the chemical reaction paths of carbon chain radicals (Cernicharo \& Gu$\acute{e}$lin 1996). Steep drop off in the abundance of long C$_n$H molecules have been seen in IRC +10216, for example, C$_8$H was observed to be a factor of 6-10 less abundant than C$_6$H (Cernicharo \& Gu$\acute{e}$lin 1996; Gu$\acute{e}$lin et al. 1997). Bell et al. (1999) also found a steep fall-off in abundance of longer C$_n$H chains in TMC-1, and they concluded that long C$_n$H chains were less likely to be abundant in the diffuse gas. However, C$_8$H were only detected in two molecular clouds, i.e. Lupus-1A (Sakai et al. 2010a) and TMC-1 (Bell et al. 1999). As such, it was important to observe carbon chain radicals toward more sources.

Given the astrochemical importance of carbon chain molecules, their formation mechanism is still a matter of some controversy (e.g. Knight et al. 1986; Winnerwisser \& Herbst 1987), and the study of $^{13}$C isotopic fractionation provides a way to discriminate the formation mechanism (Takano et al. 1998; Furuya et al. 2011).
Takano et al. (1998) observed the $^{13}$C substitutions of HC$_3$N in TMC-1 and found significant difference in abundance among the $^{13}$C isotopic species, which differs from those in Sgr B2, Orion KL, and IRC+10216. Based on these results and on the reaction rate coefficients, they conclude that the most important formation reaction of HC$_3$N is probably the reaction between C$_2$H$_2$ and CN. Abundance difference among $^{13}$C isotopologues of CCS, CCH, C$_3$S and C$_4$H were also found in TMC-1 (Sakai et al. 2007a, 2010b, 2013), which were believed to be caused by different formation processes or $^{13}$C isotope exchange reactions. Given that most of these studies are limited to Taurus molecular cloud, more sources are needed to investigate whether the abundance difference among $^{13}$C isotopologues of carbon-chain molecules is general.

Friesen et al. (2013) detected multiple HC$_7$N clumps within the young, cluster-forming Serpens South region in the Aquila rift. Their result extended the known star-forming regions containing significant HC$_7$N emission from typically quiescent regions, like the Taurus molecular cloud (Kroto et al. 1978), to more complex, active environment. Nakamura et al. (2014) found that CCS is extremely abundant along the main filament in Serpens South, while high CCS column density is typical of the carbon-chain producing regions suggested by Hirota et al. (2009). Detailed studies of this source will promote our understanding of the formation and destruction mechanisms of long carbon-chain molecules, thus we search for cyanopolyynes such as HC$_5$N, HC$_7$N, HC$_9$N, hydrocarbons (e.g. C$_6$H and C$_8$H), and other carbon-chain molecules of astrochemical interest with Shanghai Tian Ma Radio Telescope (TMRT) toward the strongest HC$_7$N emission clump, Serpens South 1a, which lies directly to the north of a filamentary ridge.

\section{OBSERVATIONS AND DATA REDUCTION }
\label{observation}

We performed observations of carbon-chain molecules at Ku and C band towards the Serpens South 1a in 2015 May with the TMRT. The TMRT is a new 65 m diameter fully-steerable radio telescope located in the western suburbs of Shanghai, China (Yan et al. 2015). Five cryogenically cooled receivers covering the frequency ranges 1.25-1.75 GHz (L), 2.2-2.4 GHz (S), 4.0-8.0 GHz (C) 8.2-9.0 GHz (X) and 11.5-18.5 GHz (Ku), respectively, are available now. The Digital backend system (DIBAS) of TMRT is an FPGA-based spectrometer based upon the design of Versatile GBT Astronomical Spectrometer (VEGAS) (Bussa 2012). For molecular line observations, DIBAS supports a variety of observing modes, including 19 single sub-band modes and 10 eight sub-band modes. The bandwidth of single sub-band mode varies from 1250 MHz to 11.7 MHz. The spectrometer supports 8 fully tunable sub-bands within 1300 MHz bandwidth. The bandwidth of each sub-band is 23.4 MHz for mode 20-24, while it is 15.6 MHz for mode 25-29. The center frequency of the sub-band is tunable to an accuracy of 10 KHz.

Figure 1 (upper left) shows the HC$_7$N $\emph{J}$=21-20 and NH$_3$ integrated intensity observed with the Robert C. Byrd Green Bank Telescope (GBT) overlaid as contours over the thermal continuum emission from dust at sub-millimeter wavelength (Friesen et al. 2013; Andr$\acute{e}$ et al. 2010). The 54$''$ FWHM TMRT beam at 18 GHz (yellow circle) and 121$''$ FWHM TMRT beam at 8 GHz (blue circle) are shown. The adopted coordinates for our searches were: R.A. (2000) = 18:29:57.9, DEC (2000) = -1:56:19.0. The pointing accuracy is better than 12$''$. The DIBAS mode 20 was adopted for Ku band observation, with 8 spectral windows, each of which has 4096 channel and a bandwidth of 23.4 MHz, supplying a velocity resolution of about {0.12 \kms\ (15 GHz) and 0.09 \kms\ (18 GHz) at Ku band. The DIBAS mode 22 was adopted for C band observation, with 8 spectral windows, each of which has 16384 channel and a bandwidth of 23.4 MHz, supplying a velocity resolution of about 0.05 \kms\ at 8 GHz. The intensity were calibrated by injecting periodic noise, and the accuracy of the intensity calibration is 20\%. The system temperature was about 30-80 K at Ku band, and 20-50 K at C band. We performed deep integrations to search for emission lines from HC$_9$N, C$_6$H, C$_8$H, and the $^{13}$C substitutions of HC$_3$N (H$^{13}$CCCN, HC$^{13}$CCN, and HCC$^{13}$CN), reaching rms noise levels ($\sigma$) of 3-10 mK $T_{MB}$ in a smoothed resolution of about 0.1 \kms. All the data were taken in position switching mode. Both the on-source and off-source time are 3 minutes. The resulting antenna temperatures were scaled to main beam temperatures ($T_{MB}$) by using a main beam efficiency of 0.6 for a moving position of sub-reflector at Ku band, and a main beam efficiency of 0.6 at C band (Wang et al. 2015). Table 1 lists parameters of the observed lines, including rest frequency, upper energy level E$_u$, FWHM beam and rms noise in emission free region.
The rest frequencies and their uncertainties of observed lines are obtained from the molecular database at ``Splatalogue''\footnote{\tt www.splatalogue.net.}, which is a compilation of the Jet Propulsion Laboratory (JPL, Pickett et al. 1998), Cologne Database for Molecular Spectroscopy catalogues (CDMS, M$\ddot{u}$ller et al. 2005), and Lovas/NIST catalogues (Lovas 2004).

The data processing was conducted using \textbf{GILDAS} software package\footnote{\tt http://www.iram.fr/IRAMFR/GILDAS.}, including CLASS and GREG. Linear baseline subtractions were used for all the spectra. For each transition, the spectra of subscans, including two polarizations, were averaged to reduce rms noise levels.

\section{OBSERVATIONAL RESULTS}
\label{result}

We detected five transitions of HC$_9$N, four transitions of HC$_7$N, and three transitions of HC$_5$N at C and Ku band. The hyperfine structure of HC$_5$N $\emph{J}$=6-5 $\emph{F}$=5-4 at 15974.9336 MHz was resolved for the first time. Ohishi \& Kaifu (1998) detected HC$_5$N $\emph{J}$=6-5 in TMC-1 with Nobeyama 45m radio telescope. The spectral resolution was 37 kHz ($\sim$0.7 \kms\ at 15 GHz) in their observation. The separation between HC$_5$N $\emph{J}$=6-5 $\emph{F}$=5-4 and HC$_5$N $\emph{J}$=6-5 is 32 kHz, thus the hyperfine structure was not resolved at the time. Two hyperfine components of HC$_9$N ($\emph{J}$=13-12 $\emph{F}$=12-11 at 7553.463 MHz and HC$_9$N $\emph{J}$=13-12 $\emph{F}$=14-13 at 7553.474) were detected. Spectra line profiles of HC$_5$N $\emph{J}$=6-5, HC$_7$N $\emph{J}$=15-14 and HC$_9$N $\emph{J}$=31-30 are also shown in Figure 1.

C$_3$S 3-2 was detected with a peak intensity of 0.46 K. Spectra line profile of C$_3$S $\emph{J}$=3-2 are shown in Figure 2 (upper panel).

The four fine and hyperfine structure components of the $\emph{J}$=13/2-11/2 transition of C$_6$H were detected with a peak intensity of 52-76 mK.
The doublet lines with partially resolved hyperfine structure of $\emph{J}$=31/2-29/2 lines of C$_8$H were also detected. The intensity ratio of C$_6$H to C$_8$H is about 6:1. Spectra line profiles of the $\emph{J}$=13/2-11/2 transitions of C$_6$H $^2\prod_{3/2}$ and the $\emph{J}$=31/2-29/2 transitions of C$_8$H $^2\prod_{3/2}$ are shown in Figure 2 (lower panel).

The $^{13}$C substitutions of HC$_3$N (H$^{13}$CCCN, HC$^{13}$CCN, and HCC$^{13}$CN) $\emph{J}$=2-1 were also detected. Spectra line profiles of HC$_3$N and its $^{13}$C isotopomers are shown in Figure 3. The intensity of HCC$^{13}$CN is stronger than those of HC$^{13}$CCN and H$^{13}$CCCN, with the difference in the peak intensities at a level of about 4$\sigma$ ($\sigma \sim$4 mK). On the other hand, there is no significant difference in intensity between the HC$^{13}$CCN and H$^{13}$CCCN lines. This is similar to observations of TMC-1 (Takano et al. 1998). This result indicates that the abundance difference between H$^{13}$CCCN and HC$^{13}$CCN, and HCC$^{13}$CN is not specific to TMC-1, but seems to be common for the carbon-chain-rich clouds. H$^{13}$CCCN $\emph{J}$=2-1, HC$^{13}$CCN $\emph{J}$=2-1 and HCC$^{13}$CN $\emph{J}$=2-1 have been observed in TMC-1 with Nobeyama 45m telescope (Ohishi \& Kaifu 1998; Takano et al. 1998; Kaifu et al. 2004). Limited by their spectral resolution (37 kHz, $\sim$0.6 \kms\ at 18 GHz), the hyperfine components of these transitions were partially resolved. Resolved hyperfine components including the weak satellite components are detected for the first time.

The line parameters are summarized in Table 2, including peak temperature ($T_{MB}$), FWHM linewidth, centroid velocity, integrated line intensity ($\int T_{MB}d\nu$), and column density. These parameters and their errors are derived by fitting the Gaussian profile to each spectrum. Because the signal to noise ratio is not high enough, it is hard to fit the weaker hyperfine component ($^2\prod_{3/2}$ 31/2-29/2 e) of C$_8$H well. We could see from Table 2 that centroid velocities and FWHM linewidths of these carbon-chain molecule lines are similar, which are around 7.6 km s$^{-1}$ and 0.6 km s$^{-1}$, respectively, suggesting that these emission come from similar region. The linewidths are consistent with GBT observations of HC$_7$N in this source (0.56 km s$^{-1}$; Friesen et al. 2013). The linewidths are significantly higher than those in TMC-1 and Lupus-1A, which are reported to be 0.2 $\sim$ 0.3 km s$^{-1}$ (Li et al. 2010; Sakai et al. 2010a), implying that these emission come from more complex and active regions than other two sources.

\section{DISCUSSIONS}
\label{discussion}

\subsection{Column densities of Carbon-chain molecules}
\label{abundance}

As shown in Figure 1, the size of HC$_7$N clump Serpens South 1a is larger than the FWHM TMRT beam at Ku band, but smaller than the FWHM TMRT beam at C band. Thus it is inappropriate to estimate the rotational temperature and column density with rotational diagrams. Transitions in the 16-18 GHz range were used to derive the column densities of each molecule, and lines that were used to derive column densities were labeled with 'a' in Table 2. According to Figure 1, the size of the HC$_7$N clump Serpens South 1a is about 2 arcmin, therefore, the beam filling factors is $\sim$1. By assuming local thermodynamic equilibrium (LTE) conditions and optically thin for all the molecules, the column densities of these species are estimated with the following equation (Cummins et al. 1986):
\begin{equation}
N=\frac{3kW}{8 \pi^3 \nu S \mu^2}(\frac{T_{ex}}{T_{ex}-T_{bg}})\times Q(T_{ex})exp(E_u/kT_{ex})
\end{equation}
where $k$ is the Boltzmann constant in erg K$^{-1}$, $W$ is the observed line integrated intensity in K Km s$^{-1}$, $\nu$ is the frequency of the transition in Hz, and $S\mu^2$ is the product of the total torsion-rotational line strength and the square of the electric dipole moment. $T_{ex}$ and $T_{bg}$ (=2.73 K) are the excitation temperature and background brightness temperature, respectively. $E_u/k$ is the upper level energy in K, and $Q(T_{ex})$ is the partition function. Values of $E_u/k$ and $S\mu^2$ are taken from the ``SPLATALOGUE'' spectral line catalogs. Friesen et al. (2013) derived a kinetic temperature of 10.8 K with observation of NH$_3$. Usually, kinetic temperature is larger than excitation temperature in starless cores. We adopt an excitation temperature of 7 K, which is similar to those in TMC-1 (Bell et al. 1998; Sakai et al. 2007a) and Lupus-1A (Sakai et al. 2010a). The partition function $Q(T_{ex})$ of each molecule are estimated by fitting the partition function at different temperatures given in CDMS. For most of our molecules, increasing the excitation temperature by 2 K would increase or decrease the resulting column density by fewer than 10\%. The maximum variation comes from HC$_9$N, in which an increase in excitation temperature of 2 K would decrease column density by 30\%. The derived column densities and their errors are listed in Table 2. The errors come from the uncertainties of integrated intensities, which are obtained by Gaussian fitting. Decreasing the excitation temperature by 2 K would increase or decrease the resulting column density by fewer than 10\% for most of our molecules. The maximum variations come from HC$_7$N and HC$_9$N, in which a decrease in excitation temperature of 2 K would increase column density by 32\% and 110\%, respectively.

The column densities of these carbon-chain molecules range from $10^{12}$ to $10^{13}$ cm$^{-1}$. Figure 4 shows comparison of the column densities between Serpens South 1a, TMC-1 and Lupus-1A. We could see from Figure 4 that the column densities of long carbon-chain molecules in Serpens South 1a are comparable to those in TMC-1 and Lupus-1A (Sakai et al. 2010a). Thus we could say that the long carbon-chain molecules in Serpens South 1a are as abundant as in TMC-1. Therefore, like Lupus-1A, Serpens South 1a could also be regarded as a ``TMC-1 like source''. This provides further evidence that the extraordinary richness of carbon-chain molecules in TMC-1 should not be ascribed to some special regional reasons, but should be considered as a more general phenomenon.

The abundance ratio for HC$_7$N: HC$_9$N is (2.0$\pm$0.2):1, which is similar to the abundance ratios found for TMC-1 and Lupus-1A, 4.8 and 2.5, respectively (Sakai et al. 2008, 2010a). The abundance ratio for C$_6$H: C$_8$H is calculated to be (1.8$\pm$0.1):1, which seems to be not as steep as the ratios observed in TMC-1 and Lupus-1A, 6.5 and 10, respectively (Br$\ddot{u}$nken et al. 2007; Sakai et al. 2008, 2010a). However, it should be noted that the column density is derived from one rotational transition of each molecule in this paper. In this case, the column densities of C$_6$H and C$_8$H (or HC$_7$N and HC$_9$N) are derived from the transitions having different upper-state energies (See Table 1). So variation of the rotation temperature will affect the column densities for the two molecules differently. In general, the rotation temperature tends to be higher for longer chains (e.g. Ohishi \& Kaifu 1998; Bell et al. 1998). As calculated above, if the rotation temperature of HC$_9$N is higher than HC$_7$N by 2 K, the abundance ratio for HC$_7$N: HC$_9$N could be higher by $\sim$~30\%. If the rotation temperature of C$_8$H is higher than C$_6$H by 2 K, the abundance ratio  could be higher by about 3\%. Our results suggest that the abundance ratio for HC$_7$N: HC$_9$N is similar to those in TMC-1 and Lupus-1A, while the abundance ratio for C$_6$H: C$_8$H is statistically significantly lower than those in TMC-1 and Lupus-1A. 

\subsection{Abundance Ratio of HC$_3$N Isophomers}
\label{temperature}

Study of $^{13}$C isotopic fractionation provides a way to discriminate the formation mechanism of carbon-chain molecules (Takano et al. 1998). Numerical calculations suggest that molecules formed from carbon atoms have carbon isotope ratios (CX/$^{13}$CX) greater than the elemental abundance ratio of [$^{12}$C/$^{13}$C], while molecules formed from CO molecules have CX/$^{13}$CX ratios smaller than the [$^{12}$C/$^{13}$C] (Furuya et al. 2011). We investigate the $^{13}$C isotopic fractionation with observations of $^{13}$C isotopologues of HC$_3$N in Serpens South 1a.

The HC$_3$N 2-1 transition has six hyperfine components, all of which were clearly seen in our spectra.
We calculate the optical depth of HC$_3$N using the ``hfs'' fitting method (McGee et al. 1997)
in GILDAS CLASS package and found that it is optically thin, with optical depth smaller than 0.1, see Figure 3.
This is obviously smaller than in TMC-1, in which an optical depth of 0.47 was obtained from
HC$_3$N $\emph{J}$=2-1 transition (Li \& Goldsmith 2012), suggesting that the column density of HC$_3$N in Serpens South 1a should be lower than that in TMC-1 at the same temperature. The column densities of the normal and the $^{13}$C
isotopomers of HC$_3$N were calculated assuming the LTE using Equation 1, also
with the excitation temperature of 7 K. The integrated intensity (W) and $S\mu^2$ of the strongest hyperfine components were used to derive the column densities for the $^{13}$C isotopomers of HC$_3$N. We didnot use the strongest component to derive the column density of the normal HC$_3$N because of its larger uncertainty in comparison with other components. Lines that were used for analysis were also labeled with 'a' in Table 2. The column densities obtained are listed in Table 2.
The column density of normal HC$_3$N was comparable to that of TMC-1 ($(1.6\pm0.1)\times10^{14}$ cm$^{-2}$, Takano et al. 1998).
The ratios of the column densities of the three $^{13}$C istopic species are 1.00:(1.11$\pm$0.15):(1.47$\pm$0.18) for [H$^{13}$CCCN]:[HC$^{13}$CCN]:[HCC$^{13}$CN]. Result present here is similar to TMC-1, in which the abundance ratios are observed to be 1.0:1.0:1.4 for [H$^{13}$CCCN]:[HC$^{13}$CCN]:[HCC$^{13}$CN] at the cyanopolyyne peak of TMC-1. This result implies that the $^{13}$C isotope is also concentrated in the carbon atom adjacent to the nitrogen atom in HC$_3$N, which supports that C$_2$H$_2$+CN is probably the most important reaction in producing HC$_3$N (cf. Takano et al. 1998).

The carbon isotopic ratios ($^{12}$C/$^{13}$C) were further calculated from the obtained column densities of the normal and isotopic species of HC$_3$N. The carbon isotopic ratios obtained from H$^{13}$CCCN, HC$^{13}$CCN and HCC$^{13}$CN are 78$\pm$9, 70$\pm$8 and 53$\pm$4. The average $^{12}$C/$^{13}$C ($\sim$67$\pm$7) does not significantly deviate from the value of $\sim$ 70 in the solar neighborhood (Wilson \& Rood 1994).

The [HC$_3$N]/[H$^{13}$CCCN], [HC$_3$N]/[HC$^{13}$CCN] and [HC$_3$N]/[HCC$^{13}$CN] ratios were derived to be 79$\pm$11, 75$\pm$10 and 45$\sim$55 in TMC-1 (Takano et al. 1998). Therefore, the [HC$_3$N]/[H$^{13}$CCCN] and [HC$_3$N]/[HC$^{13}$CCN] ratio are similar for TMC-1 and Serpens South 1a. Langer \& Graedel (1989) modeled the time evolution of the $^{12}$C/$^{13}$C ratios of various species with ion-molecule chemistry of nitrogen-, oxygen-, and carbon-bearing molecules and isotopic ratios. They found that the ratios depend sensitively on physical conditions like temperature and density, implying similar physical evolutionary stages of Serpens South 1a to that of TMC-1. Observations toward more sources are needed to obtain a full understanding of the behavior of $^{13}$C species in molecular clouds.

\section{SUMMARY}
\label{summary}

We have carried out carbon-chain molecular line observations toward Serpens South 1a with the TMRT. Several carbon-chain molecules were detected, including HC$_5$N, HC$_7$N, HC$_9$N, C$_3$S, C$_6$H, C$_8$H, and $^{13}$C substitutions of HC$_3$N. For the first time, some transitions and hyperfine components from HC$_5$N, HC$_9$N, and $^{13}$C substitutions of HC$_3$N were detected or resolved.

We calculated the column densities of these molecules and found that the column densities of these carbon-chain molecules in Serpens South 
1a are comparable to those in other two carbon-chain molecule rich sources, TMC-1 and Lupus-1A. Thus this source could be regarded as 
another ``TMC-1 like source". The column density ratio of HC$_7$N: HC$_9$N is similar to those in TMC-1 and Lupus-1A, while 
the abundance ratio of C$_6$H: C$_8$H seems to be statistically significantly lower than those in TMC-1 and Lupus-1A. 

We also derived the abundance ratios of HC$_3$N isophomers, and made comparison with TMC-1. We found an average $^{12}$C/$^{13}$C ratio of about 67$\pm$7, which does not significantly deviate from the value in the solar neighborhood. There is no difference for carbon isotopic ratios between TMC-1 and Serpens South 1a, reflecting similar physical evolutionary conditions of these two sources.

\acknowledgments We would like to thank the anonymous referee for his/her very constructive suggestions that led to significant improvements in the paper. We also thank the assistance of the TMRT operators during the observations. J.L. wish to thank Y. Gong for help with calculation of column densities. This work is partly supported by China Ministry of Science and Technology under State Key Development Program for Basic Research (2012CB821800), and partly supported by the Natural Science Foundation of China under grants of 11103006.

%

\begin{table}
\scriptsize
    \begin{center}
      \caption{Observed Transitions and Telescope Parameters}\label{tab:source}
      \begin{tabular}{lcccccccc}
      \\
    \hline
    \hline
(1) Species & (2) Transitions & (3) Rest Freq.   & (4) $E_{u}$  &  (5)  Observing date  & (6) FWHM beam      & (7) rms         \\
            &                 &   (MHz)          &     (K)      &                  & ($''$)     &  (mK )      \\
\hline
HC$_5$N & $\emph{J}$=3-2 $\emph{F}$=2-1   & 7987.779(0.005)      &    0.8       & 2015June23  &     121    &  10       \\
        & $\emph{J}$=3-2 $\emph{F}$=3-2   & 7987.991(0.005)      &    0.8       &    &    121    &   10     \\
        & $\emph{J}$=3-2 $\emph{F}$=4-3   & 7988.041(0.005)      &    0.8       &   &    121    &    10      \\
        & $\emph{J}$=5-4 $\emph{F}$=4-3   & 13313.261(0.002)     &    1.9       & 2015May8  &    73     &    16    \\
        & $\emph{J}$=5-4 $\emph{F}$=5-4   & 13313.312(0.002)     &    1.9       &   &    73      &   16     \\
        & $\emph{J}$=5-4 $\emph{F}$=6-5   & 13313.334(0.002)     &    1.9       &   &    73        &  16       \\
        & *$\emph{J}$=6-5 $\emph{F}$=5-4  & 15975.9336(0.0001)    &    2.7       & 2015May16  &    61          &  45       \\
        & $\emph{J}$=6-5         & 15975.966(0.001)     &    2.7       &   &    61          &  45      \\
\hline
HC$_7$N & $\emph{J}$=7-6 $\emph{F}$=6-5   & 7895.989(0.002)     &    1.5        & 2015June23 &    123      &    5      \\
        & $\emph{J}$=7-6 $\emph{F}$=7-6   & 7896.010(0.02)     &    1.5        &  &    123       &   5           \\
        & $\emph{J}$=7-6 $\emph{F}$=8-7   & 7896.023(0.002)     &    1.5        &  &    123       &   5        \\
        & $\emph{J}$=12-11       & 13535.998(0.001)    &    4.2        & 2015May8  &    72        &   14        \\
        & $\emph{J}$=13-12       & 14663.993(0.001)    &    4.9        & 2015May5 &    66        &   19        \\
        & $\emph{J}$=15-14       & 16919.979(0.001)    &    6.5        & 2015May16 &    57        &   34    \\
\hline
HC$_9$N &*$\emph{J}$=13-12 $\emph{F}$=12-11 & 7553.462(0.002)   &    2.5        & 2015June23 &    129     &    3    \\
        & *$\emph{J}$=13-12 $\emph{F}$=14-13 & 7553.474(0.002)   &    2.5        &  &    129      &   3    \\
        & $\emph{J}$=24-23       & 13944.832(0.001)    &    8.4        & 2015May16 &    70      &    6    \\
        & $\emph{J}$=28-27       & 16268.950(0.01)    &    11.3       & 2015June11 &    60        &  11      \\
        & $\emph{J}$=29-28       & 16849.979(0.001)    &    12.1       & 2015June11 &    58       &   4    \\
        & $\emph{J}$=31-30       & 18012.033(0.001)    &    13.8       & 2015May25 &    54       &   12    \\
             \hline
C$_3$S  & $\emph{J}$=3-2         & 17342.256(0.001)    &    1.7        & 2015May5 &    56          & 51     \\
\hline
C$_6$H  &$^2\prod_{3/2}$ $\emph{J}$=13/2-11/2 $\emph{F}$=7-6 e & 18020.574(0.005) &  3.0   &  2015May22,25     &   54  &      6        \\
        &$^2\prod_{3/2}$ $\emph{J}$=13/2-11/2 $\emph{F}$=6-5 e & 18020.644(0.005) &  3.0    &      &   54  &      6            \\
        &$^2\prod_{3/2}$ $\emph{J}$=13/2-11/2 $\emph{F}$=7-6 f & 18021.752(0.005) &  3.0    &      &   54  &      6           \\
        &$^2\prod_{3/2}$ $\emph{J}$=13/2-11/2 $\emph{F}$=6-5 f & 18021.818(0.005) &  3.0    &      &   54  &      6             \\
\hline
C$_8$H  &$^2\prod_{3/2}$ 31/2-29/2 e         & 18186.652(0.003) &  7.1   &  2015May22,25,26     &   53  &      5           \\
        &$^2\prod_{3/2}$ 31/2-29/2 f         & 18186.782(0.003) &  7.1   &      &   53  &      5              \\
\hline
 HC$_3$N    &  $\emph{J}$=2-1 $\emph{F}$=2-2  & 18194.9206(0.0008) &   1.3    &  2015May22,25,26    &   53    &   5   \\
            &  $\emph{J}$=2-1 $\emph{F}$=1-0  & 18195.1364(0.0006)  &   1.3    &      &   53    &   5     \\
            &  $\emph{J}$=2-1 $\emph{F}$=2-1  & 18196.2183(0.0005) &  1.3     &      &   53     &   5     \\
            &  $\emph{J}$=2-1 $\emph{F}$=3-2  & 18196.3119(0.0007) &  1.3     &        &   53    &   5     \\
            &  $\emph{J}$=2-1 $\emph{F}$=1-2  & 18197.078(0.001)  &   1.3    &        &   53     &   5      \\
            &  $\emph{J}$=2-1 $\emph{F}$=1-1  & 18198.3756(0.0009) &   1.3    &        &   53    &     5    \\
H$^{13}$CCCN &  *$\emph{J}$=2-1 $\emph{F}$=1-0 & 17632.6699(0.005) &   1.3    &  2015May22,25,26       &   55   &   3  \\
            &  $\emph{J}$=2-1 $\emph{F}$=2-2  & 17632.685(0.007)  &   1.3    &         &   55   &    3    \\
            &  *$\emph{J}$=2-1 $\emph{F}$=2-1  & 17633.7506(0.0005) &   1.3    &          &  55      &  3       \\
            &  $\emph{J}$=2-1 $\emph{F}$=3-2  & 17633.844(0.004)  &   1.3  &          &  55      &   3     \\
            &  *$\emph{J}$=2-1 $\emph{F}$=1-1  & 17635.9084(0.0005) &   1.3  &          &  55       &   3   \\
HC$^{13}$CCN & *$\emph{J}$=2-1 $\emph{F}$=2-2 & 18117.727(0.0006) &    1.3  &   2015May22,25,26       &  54      &    3  \\
            & *$\emph{J}$=2-1 $\emph{F}$=1-0  & 18117.9441(0.0006) &    1.3  &          &  54      &    3    \\
            & $\emph{J}$=2-1 $\emph{F}$=2-1   & 18119.029(0.005) &    1.3  &          &   54     &    3    \\
            & $\emph{J}$=2-1 $\emph{F}$=3-2   & 18119.122(0.005)  &   1.3  &           &   54     &   3    \\
            & *$\emph{J}$=2-1 $\emph{F}$=1-1   & 18121.1825(0.0006)  &   1.3  &           &   54     &   3    \\
HCC$^{13}$CN & *$\emph{J}$=2-1 $\emph{F}$=1-0  & 18119.6931(0.0005)  &  1.3   &          &   54    &   3  \\
             & $\emph{J}$=2-1 $\emph{F}$=2-1   & 18120.773(0.002) &   1.3   &  2015May22,25,26        &   54     &  3  \\
             & $\emph{J}$=2-1 $\emph{F}$=3-2   & 18120.865(0.002) &   1.3   &          &   54     &  3       \\
            & *$\emph{J}$=2-1 $\emph{F}$=1-1   & 18122.9315(0.0005) &   1.3  &           &   54     &   3    \\
\hline
      \end{tabular}
  \end{center}
  Notes.- $^*$: Lines detected for the first time. Col. (1): molecule name; Col. (2): transition, Col. (3): rest frequency; Col. (4): E$_u$; Col. (5): observing date; Col. (6): FWHM beam; Col. (7): rms noise in emission free channel, which are obtained from Gaussian fitting.
\end{table}

\begin{table}
\scriptsize
    \begin{center}
      \caption{Line Parameters of Long Carbon-chain Molecules Detected in Serpens South 1a}\label{tab:source}
      \begin{tabular}{lcccccccc}
      \\
    \hline
    \hline
(1) Species & (2) Transitions & (3) $T_{MB}$  & (4) $\triangle V$   & (5) $V_{LSR}$  & (6) $\int T_{MB}d\nu$ & (7)   $N$   \\
            &             &      (mK)  & (km s$^{-1}$)&  (km s$^{-1}$) &  (mK km s$^{-1}$) &    (cm$^{-2}$)    \\
\hline
HC$_5$N & $\emph{J}$=3-2 $\emph{F}$=2-1   & 91(9)   & 0.53(.05) & 7.63(.02) & 51(4)   &          \\
        & $\emph{J}$=3-2 $\emph{F}$=3-2   & 129(9)  & 0.59(.05) & ...      & 80(5)     &         \\
        & $\emph{J}$=3-2 $\emph{F}$=4-3   &  184(9)     &  0.56(.03) &  ...    &  109(5)  &      \\
        & $\emph{J}$=5-4 $\emph{F}$=4-3   &  243(13)  & 0.50(.13)&  ...       & 130(12)    &       \\
        & $\emph{J}$=5-4 $\emph{F}$=5-4   &  312(13)  & 0.57(.13)&  ...       & 190(12)    &       \\
        & $\emph{J}$=5-4 $\emph{F}$=6-5   &  348(13)  & 0.41(.13)& 7.54(.13)  & 151(12)   &         \\
        & *$\emph{J}$=6-5 $\emph{F}$=5-4  &   302(36)      & 0.43(.04)& ...          &  138(12)      &   \\
        & $^a\emph{J}$=6-5         &   525(36)       & 0.69(.03)&  7.51(.01)   &  385(14)   &   $1.2(.1)\times10^{13}$     \\
\hline
HC$_7$N & $\emph{J}$=7-6 $\emph{F}$=6-5   &  18(3)         & 0.56(.13) & ...   & 11(6)    &      \\
        & $\emph{J}$=7-6 $\emph{F}$=7-6   &  29(3)     & 0.40(.09)   & ...  &  12(5)     &        \\
        & $\emph{J}$=7-6 $\emph{F}$=8-7   &  29(3)     & 0.66(.18)  &  7.55(.06) &  21(3)   &     \\
        & $\emph{J}$=12-11       &  291(10)  & 0.60(.03) & 7.62(.01)  & 186(6)  &        \\
        & $\emph{J}$=13-12       &  363(19)  & 0.55(.02) & 7.53(.01)  & 214(7)  &        \\
        & $^a\emph{J}$=15-14       &   415(38)      & 0.50(.02) & 7.54(.01)    &  220(8)    &  $6.0(.2)\times10^{12}$    \\
\hline
HC$_9$N & *$\emph{J}$=13-12 $\emph{F}$=12-11 &11(1)  & 0.31(.13)&  7.63(.04) & 4(1)  &      \\
        & *$\emph{J}$=13-12 $\emph{F}$=14-13 &10(1)  & 0.28(.19)&  ..        & 2(1)    &      \\
        & $\emph{J}$=24-23         & 52(5)   & 0.44(.04)&  7.50(.01) &  24(2)  &       \\
        & $\emph{J}$=28-27       & 57(14)     &  0.59(.12)  &  7.57(.04)  &  36(5)   &       \\
        & $\emph{J}$=29-28       &  65(3)      &   0.58(.03)  &  7.57(.01)  &  41(2)   &      \\
        & $^a\emph{J}$=31-30       &  76(7)      &   0.57(.05)  &   7.62(.02) & 46(3)    &  $3.1(.2)\times10^{12}$     \\
             \hline
C$_3$S  & $^a\emph{J}$=3-2       &   502(28)  & 0.72(.06)  &   7.60(.03)   &  385(28)     &  $1.0(.1)\times10^{13}$    \\
\hline
C$_6$H  &$^2\prod_{3/2}$ $\emph{J}$=13/2-11/2 $\emph{F}$=7-6 e  & 48(2)  &  0.66(.06)  & ...       & 34(3)    &   $1.5(.1)\times10^{12}$    \\
        &$^2\prod_{3/2}$ $\emph{J}$=13/2-11/2 $\emph{F}$=6-5 e  & 47(2)  &  0.57(.6)   & 7.64(.03)  & 28(3)    &       \\
        &$^2\prod_{3/2}$ $\emph{J}$=13/2-11/2 $\emph{F}$=7-6 f  & 41(2)  &  0.49(.06)  & ...       &  21(3)     &    \\
        &$^a$ $^2\prod_{3/2}$ $\emph{J}$=13/2-11/2 $\emph{F}$=6-5 f  & 59(2)  &  0.54(.04)  & ...       &  34(3)      &    \\
\hline
C$_8$H  &$^2\prod_{3/2}$ 31/2-29/2 e &   9(2)         &  0.38(.86)  & ...         & 4(2)      &   $8.2(.2)\times10^{11}$    \\
        &$^a$ $^2\prod_{3/2}$ 31/2-29/2 f &   12(2)         &  1.11(.29)  & 7.45(.12)   & 14(3)     &     \\
\hline
 HC$_3$N    &  $\emph{J}$=2-1 $\emph{F}$=2-2  & 557(2)        & 0.55(.19) &  ...         & 324(2)     &  $7.8(.4)\times10^{13}$     \\
            &  $^a\emph{J}$=2-1 $\emph{F}$=1-0  & 705(2)        & 0.56(.19) & ...          & 420(2)      &     \\
            &  $\emph{J}$=2-1 $\emph{F}$=2-1  & 1288(45)        & 0.59(.19) & 7.60(.19)   & 811(89)   &     \\
            &  $\emph{J}$=2-1 $\emph{F}$=3-2  & 2014(45)        & 0.60(.19) & ...        & 1280(89)     &    \\
            &  $\emph{J}$=2-1 $\emph{F}$=1-2  &  39(4)        &  0.66(.06) & ...        &  28(3)     &    \\
            &  $\emph{J}$=2-1 $\emph{F}$=1-1  &  531(2)        & 0.55(.04) &  ...       & 313(2)     &    \\
H$^{13}$CCCN &  *$\emph{J}$=2-1 $\emph{F}$=1-0 &   8(2)        & 0.76(.17) &  ...        & 7(2)     &  $1.0(0.1)\times10^{12}$    \\
            &  $\emph{J}$=2-1 $\emph{F}$=2-2  &   9(2)        & 0.54(.18) &  7.77(.07) & 6(2)      &   \\
            &  *$\emph{J}$=2-1 $\emph{F}$=2-1  &  21(2)        & 0.67(.09) &  ...        & 15(2)     &     \\
            &  $^a\emph{J}$=2-1 $\emph{F}$=3-2  &   44(2)        & 0.48(.03) &  ...        & 22(1)     &    \\
            &  *$\emph{J}$=2-1 $\emph{F}$=1-1 &  12(2)       & 0.64(.19) &  ...        &  8(2)     &    \\
HC$^{13}$CCN & *$\emph{J}$=2-1 $\emph{F}$=2-2 &  15(2)        & 0.47(.12) &  ...  & 8(2)     &  $1.1(.1)\times10^{12}$    \\
            &  *$\emph{J}$=2-1 $\emph{F}$=1-0  &  14(2)        & 0.45(.14) &  ...  & 7(2)      &    \\
            &  $\emph{J}$=2-1 $\emph{F}$=2-1   &  24(2)        & 0.58(.08) &  7.65(.04)  & 14(2)     &   \\
            &  $^a\emph{J}$=2-1 $\emph{F}$=3-2   &   45(2)        & 0.53(.04) &  ...        & 25(2)     &   \\
            &  *$\emph{J}$=2-1 $\emph{F}$=1-1  &   9(2)      & 0.81(.17)   &   ...       &  7(2)   &    \\
HCC$^{13}$CN &  *$\emph{J}$=2-1 $\emph{F}$=1-0  &   20(2)        & 0.65(.13)  & ...        & 14(2)  &    \\
             &  $\emph{J}$=2-1 $\emph{F}$=2-1   &  32(2)        & 0.46(.06) &   7.56(.03) & 15(2)   &   $1.5(.1)\times10^{12}$  \\
             &  $^a\emph{J}$=2-1 $\emph{F}$=3-2   &  62(2)        & 0.49(.03) &   ...       & 33(2)   &       \\
             & *$\emph{J}$=2-1 $\emph{F}$=1-1   &  12(2)        & 0.50(.58) &  ...        & 6(4)    &       \\
\hline
      \end{tabular}
  \end{center}
  Notes.- The numbers in parentheses represent the errors in units of the last significant digits. The error of column densities are 1 standard deviation. *: Lines detected for the first time. $^a$: Lines used to derive the column density. Col. (1): molecule name; Col. (2): transition; Col. (3): peak temperature; Col. (4): linewidth; Col. (5): centroid velocity; Col. (6): integrated intensity; Col. (7): column density.
\end{table}

\begin{figure}[]
\begin{center}
\includegraphics[width=3.0in]{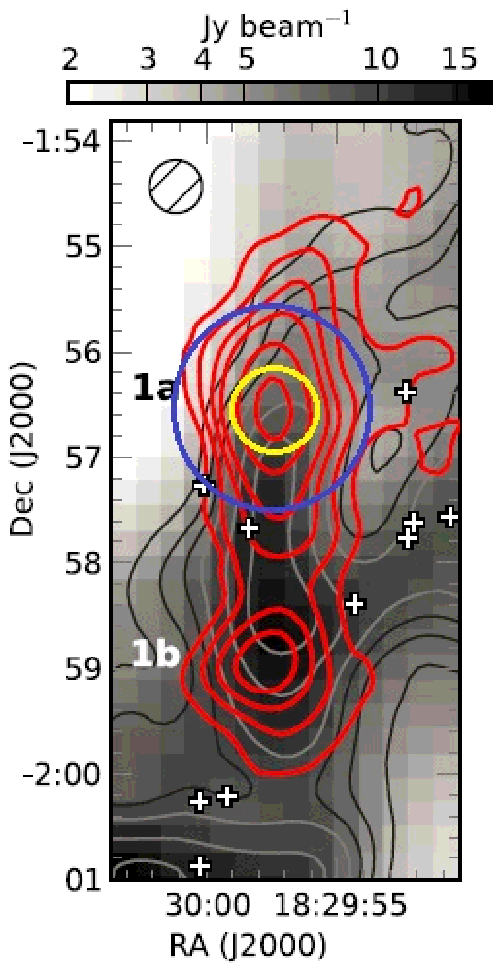}
\includegraphics[width=3.0in]{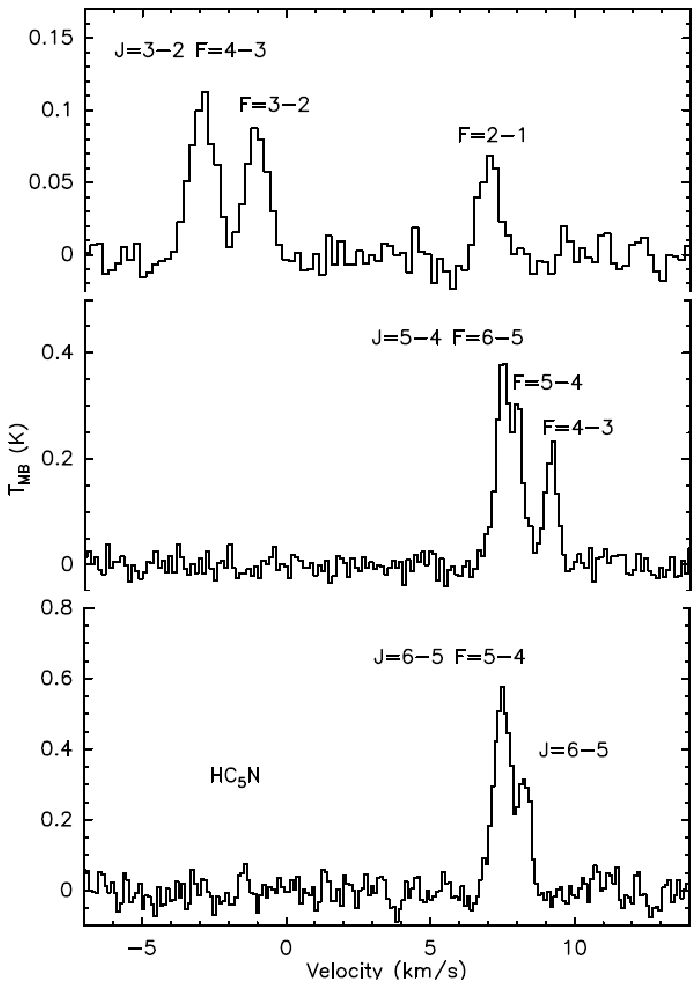}
\includegraphics[width=3.0in]{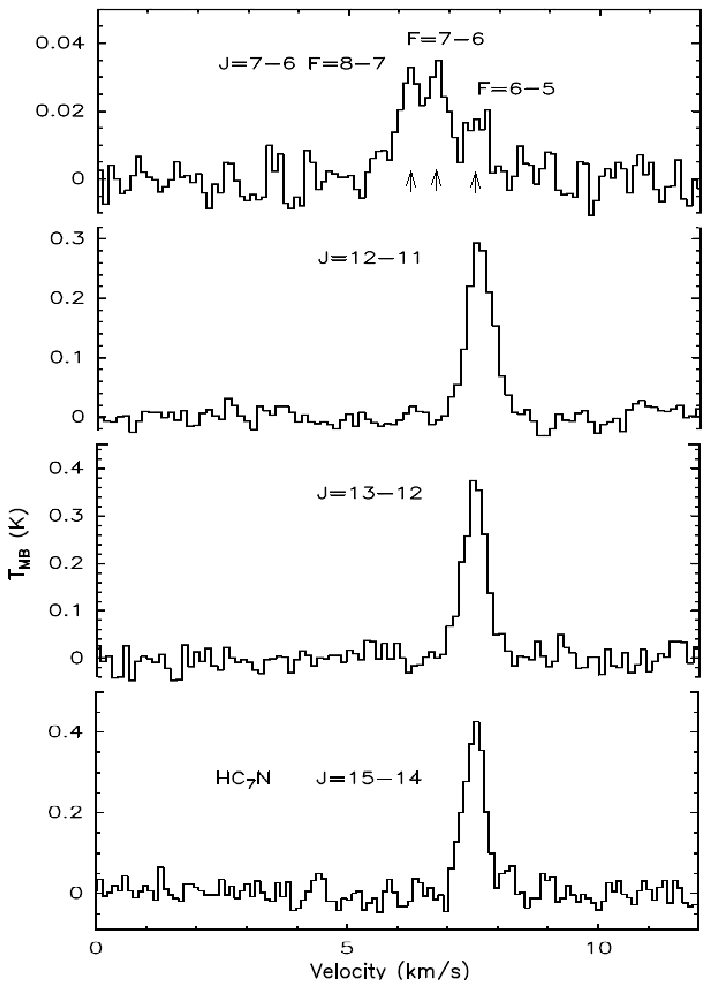}
\includegraphics[width=3.0in]{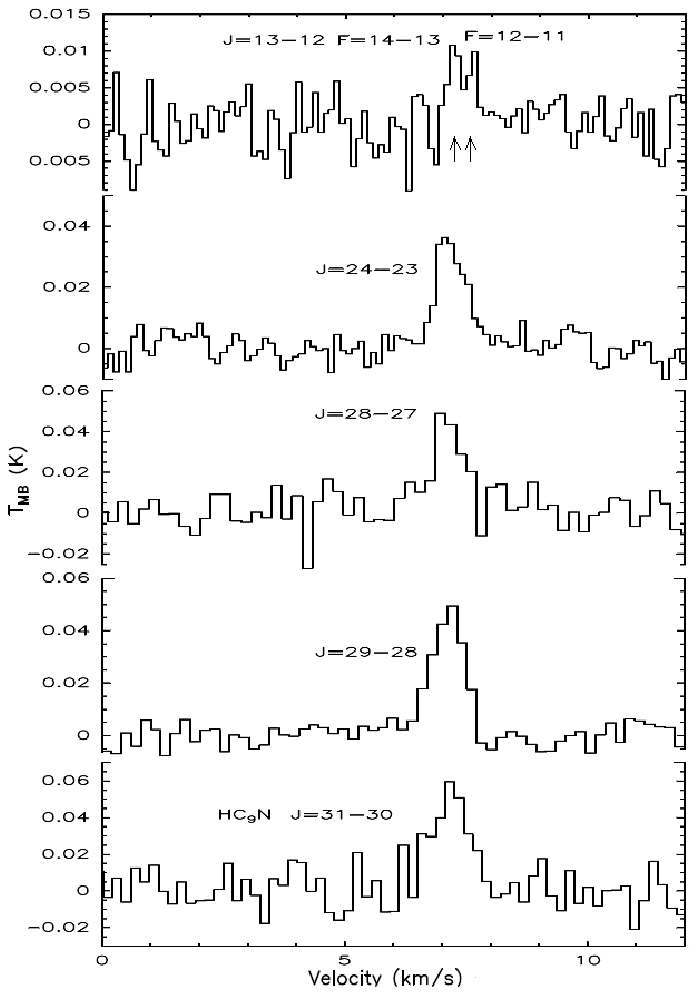}
\vspace*{-0.2 cm} \caption{\label{spec3} Upper left: Herschel SPIRE 500 $\mu$m dust continuum emission (Grey-scale; Andr$\acute{e}$ et al. 2010) towards HC$_7$N clump Serpens South 1a. Overlaid are NH$_3$ (1,1) integrated intensity contours (dark and light grey contours) and HC$_7$N 21-20 integrated intensity contours (red, at intervals of 0.15 K km s$^{-1}$). The 32$''$ FWHM GBT beam at 23 GHz is shown by the hashed circle. White crosses show Class 0 and Class I protostar locations (Friesen et al. 2013). The 54$''$ FWHM TMRT beam 18 GHz (yellow circle) and 121$''$ FWHM TMRT beam at 8 GHz (blue circle) are shown. Upper right: Spectra line profiles of HC$_5$N in Serpens South 1a. The systemic velocity is 7.6 km s$^{-1}$. Lower left: Spectra line profiles of HC$_7$N. Lower right: Spectra line profiles of HC$_9$N. Positions of the hyperfine components of HC$_7$N and HC$_9$N are indicated with arrows. }
\end{center}
\end{figure}

\begin{figure}[]
\begin{center}
\includegraphics[width=3.9in]{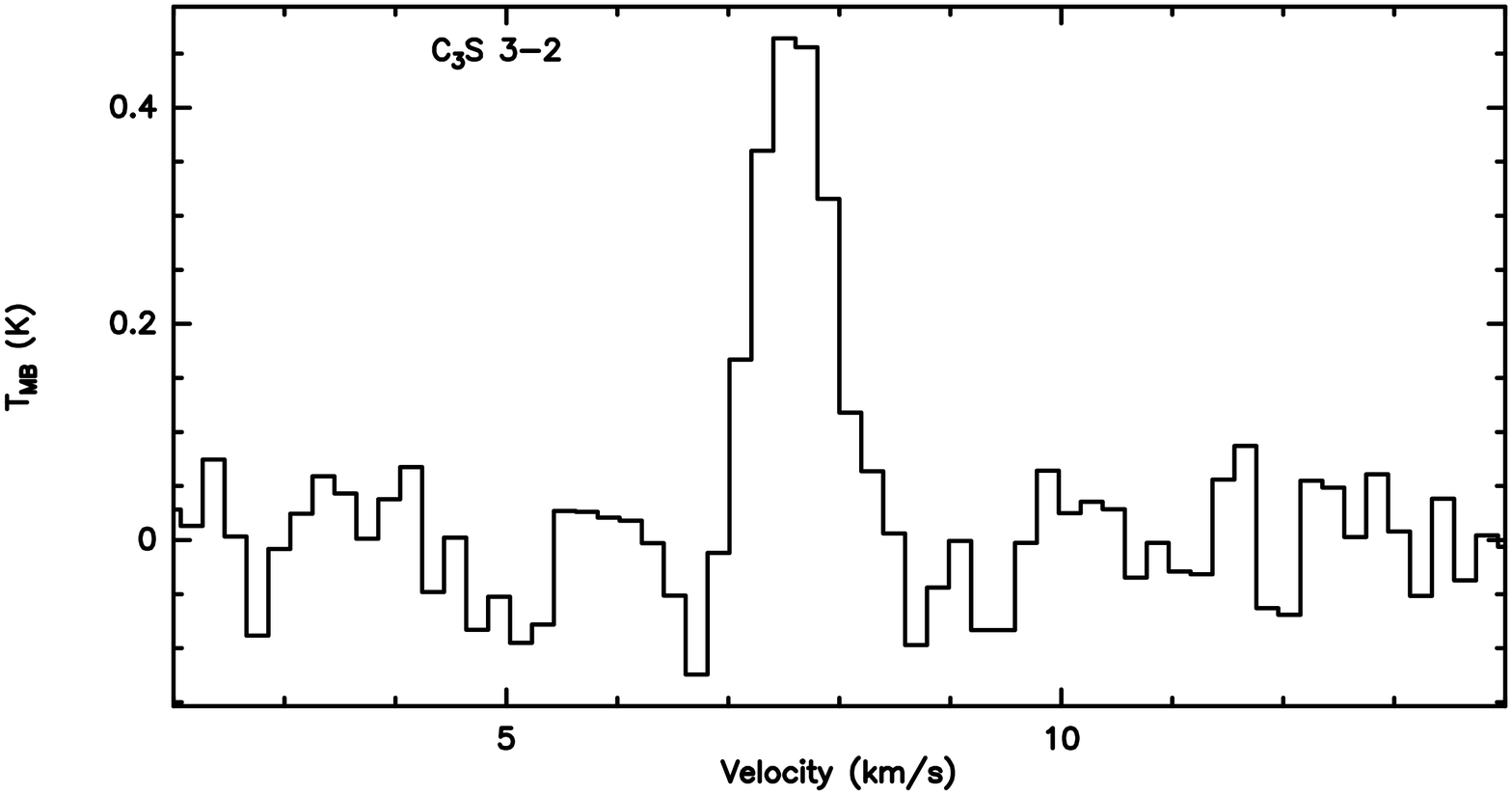}
\includegraphics[width=4.0in]{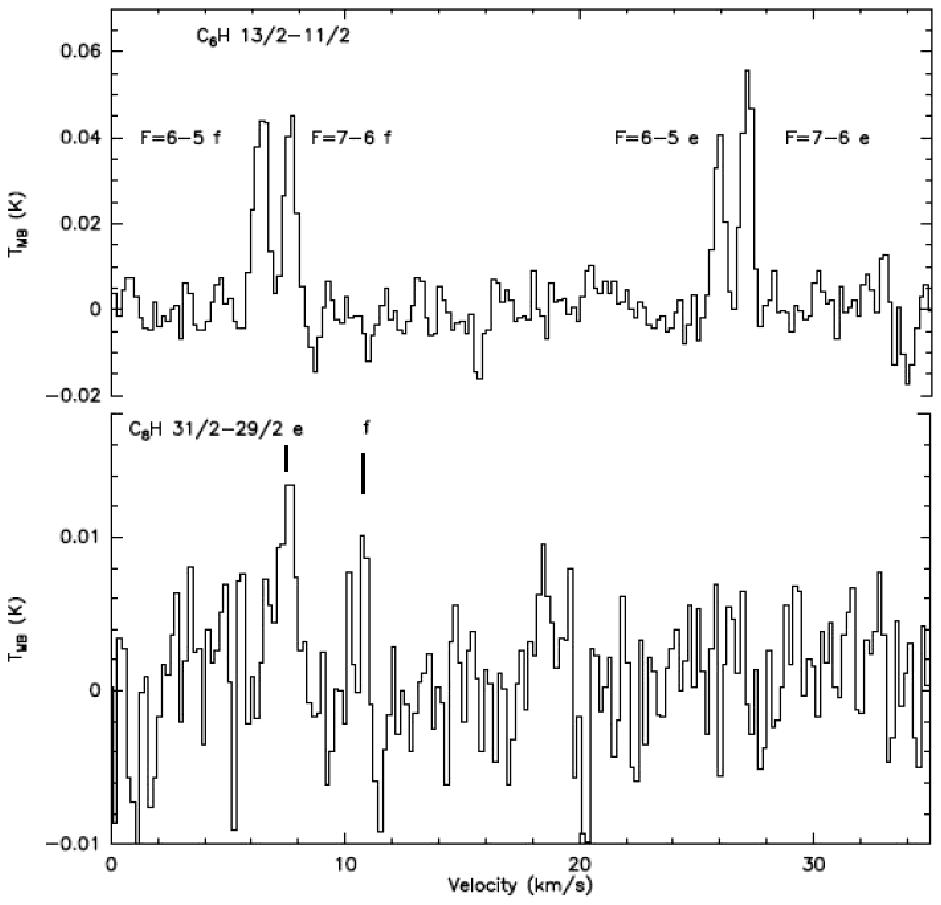}
\vspace*{-0.2 cm} \caption{\label{spec3} Upper panel: Spectra line profiles of C$_3$S in Serpens South 1a. Lower panel: Spectra line profiles of C$_6$H and C$_8$H. Both show hyperfine structures due to the spin of the hydrogen nucleus. }
\end{center}
\end{figure}

\begin{figure}[]
\begin{center}
\includegraphics[width=6.4in]{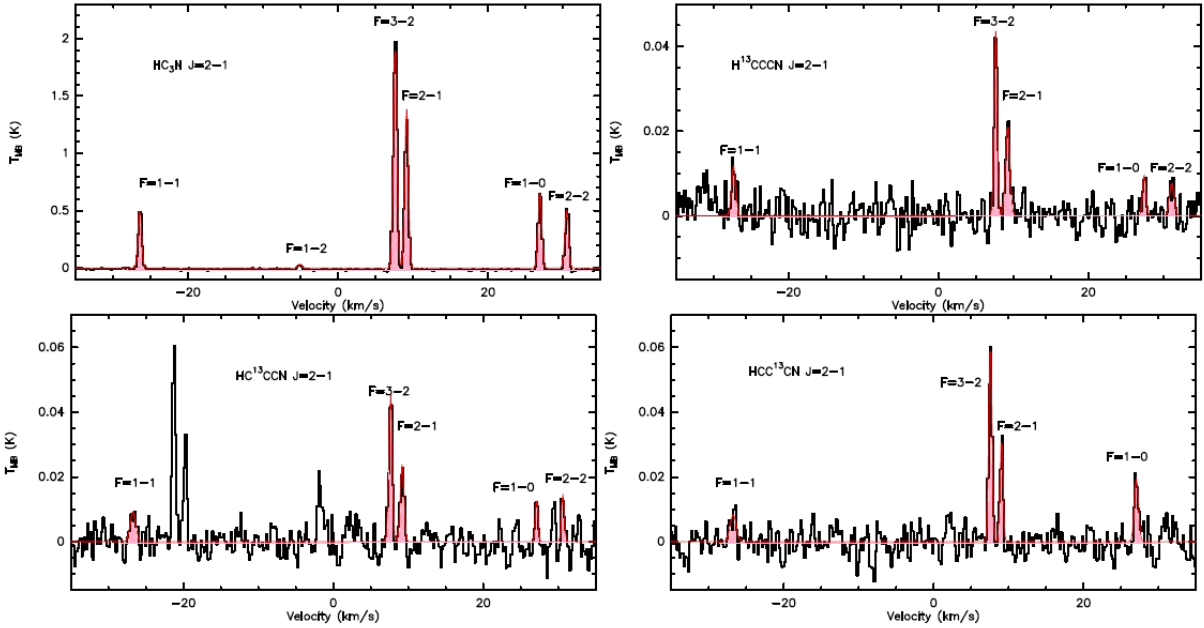}
\vspace*{-0.2 cm} \caption{\label{spec3} Spectra line profiles of HC$_3$N, H$^{13}$CCCN, HC$^{13}$CCN and HCC$^{13}$CN. All the lines show hyperfine structures. The red lines represent the hyperfine fitting using CLASS. All the hyperfine components detected are filled with pink.}
\end{center}
\end{figure}

\begin{figure}[]
\begin{center}
\includegraphics[width=3.9in]{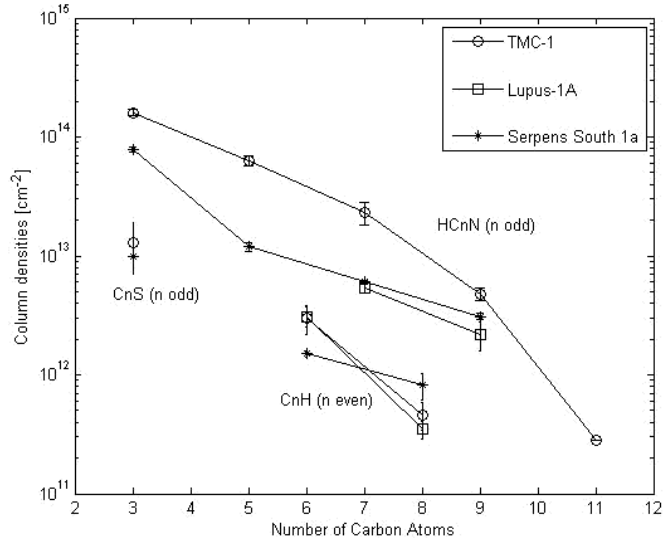}
\vspace*{-0.2 cm} \caption{\label{spec3} Comparison of the column densities between Serpens South 1a, TMC-1 and Lupus-1A. The 1$\sigma$ errors of column densities are also shown. The column densities of Serpens South 1a are comparable to those of TMC-1 and Lupus-1A, thus Serpens South 1a could be regarded as another ``TMC-1 like source''. The column densities of HC$_3$N, HC$_5$N and HC$_{11}$N in TMC-1 are taken from Takano et al. (1998), Takano et al. (1990) and Bell et al. (1997), respectively. The column densities of C$_6$H and C$_8$H in TMC-1 are taken from Sakai et al. (2007b) and Br$\ddot{u}$nken et al. (2007), respectively. The column densities of C$_3$S in TMC-1 are taken from Yamamoto et al. (1987). The column densities of carbon-chain molecules in Lupus-1A are taken from Sakai et al. (2010a). }
\end{center}
\end{figure}

\end{document}